\documentclass[ArXiv,AXISWP,accept,moreauthors,pdftex,10pt,letterpaper]{Definitions/axis} 

\firstpage{1} 
\pubvolume{xx}
\issuenum{1}
\articlenumber{5}
\pubyear{2023}
\copyrightyear{2023}

\usepackage[normalem]{ulem}

\Title{Prospects for AGN Studies with AXIS: AGN Fueling – Resolving Hot Gas inside Bondi Radius of SMBHs}

\Author{Ka-Wah Wong $^{1}$, Helen R. Russell $^{2}$, Jimmy A. Irwin $^{3}$, Nico Cappelluti $^{4}$ and Adi Foord $^{5}$}

\AuthorNames{Ka-Wah Wong, Helen Russell, Jimmy Irwin, Nico Cappelluti and Adi Foord}

\address{%
$^{1}$ \quad Department of Physics, SUNY Brockport, Brockport, NY, 14420, USA; kwong@brockport.edu\\
$^{2}$ \quad School of Physics \& Astronomy, University of Nottingham, University Park, Nottingham NG7 2RD, UK\\
$^{3}$ \quad Department of Physics and Astronomy, The University of Alabama, Tuscaloosa, AL 35487, USA\\
$^{4}$ \quad Department of Physics, University of Miami, Coral Gables, FL 33124, USA\\
$^{5}$ \quad Department of Physics, University of Maryland Baltimore County, 1000 Hilltop Cir, Baltimore, MD 21250, USA}

\abstract{Hot gas around a supermassive black hole (SMBH) should be captured within the gravitational ``sphere of influence’’, characterized by the Bondi radius.  Deep \textit{Chandra} observations have spatially resolved the Bondi radii of five nearby SMBHs that are believed to be accreting in hot accretion mode.  Contrary to earlier hot accretion models that predicted a steep temperature increase within the Bondi radius, none of the resolved temperature profiles exhibit such an increase.  The temperature inside the Bondi radius appears to be complex, indicative of a multi-temperature phase of hot gas with a cooler component at about 0.2--0.3\,keV.  The density profiles within the Bondi regions are shallow, suggesting the presence of strong outflows.  These findings might be explained by recent realistic numerical simulations that suggest that large-scale accretion inside the Bondi radius can be chaotic, with cooler gas raining down in some directions and hotter gas outflowing in others.  With an angular resolution similar to \textit{Chandra} and a significantly larger collecting area, AXIS will collect enough photons to map the emerging accretion flow within and around the ``sphere of influence’’ of a large sample of active galactic nuclei (AGNs).  AXIS will reveal transitions in the inflow that ultimately fuels the AGN, as well as outflows that provide feedback to the environment.
\emph{This White Paper is part of a series commissioned for the AXIS Probe Concept Mission; additional AXIS White Papers can be found at the  \href{http://axis.astro.umd.edu/}{AXIS website}}
.}

\begin{document}

\section{Introduction}
\label{sec:intro}

Supermassive black holes (SMBHs) are ubiquitous in the centers of galaxies \citep{Ho08}.  Accretion processes determine the growth of SMBHs and feedback mechanisms play a critical role in the evolution of their host galaxies \citep{Fab12,MN07}.  Understanding how SMBHs are fueled from the surroundings and how matter and energy are transported back to the environment is important for understanding the cosmic ecosystem.

Immersed in the diffused hot gas of the host galaxy, the gravitational potential of the black hole dominates the thermal energy of the hot gas inside the “sphere of influence,” which is defined by the characteristic Bondi radius, $R_B=2GM_{\rm BH}/c_s^2$, where $M_{\rm BH}$ is the mass of the black hole and $c_s$ is the sound speed of the gas at a distance far away from the black hole \citep{Bon52}.  Although the Bondi accretion model is likely an oversimplification of actual accretion processes, it provides a baseline for theoretical and observational comparisons of different accretion models.  Moreover, the inferred Bondi accretion rate is of great interest for a wide range of applications, such as the possible correlation between Bondi accretion rate and AGN feedback power \citep[e.g.,][]{ADA+06, RME+13}, and the accretion rate and feedback efficiency used for cosmological simulations \citep[e.g.,][]{Pil+18}.  More importantly, direct study of the hot gas within and around the Bondi “sphere of influence” is crucial for understanding the accretion processes that begin when the gas begins to be captured by the gravitational pull of the black hole \citep[see, e.g.,][]{BM99,QN00,Bag+03,Gar+10,Blandford2022,Lalakos2022,Guo2023,Olivares2023}.

Unlike the most powerful AGNs or quasars radiating at about 10\% Eddington luminosity ($L_{\rm Edd}$), which are very rare in the SMBH population, most of the SMBHs are low-luminosity AGNs (LLAGNs) radiating at $\ll L_{\rm Edd}$.  At most about a few percent of the SMBHs are shining at $\sim 10^{-5} L_{\rm Edd}$, while most of the SMBHs are much quieter ($\sim 10^{-8} L_{\rm Edd}$) \citep{Ho08}.  It is widely believed that those AGNs radiating at $\sim$$10\% L_{\rm Edd}$ are accreting close to the Eddington rate of ${\dot M}_{\rm Edd}$ in a cold accretion mode, with the classical optically thick and geometrically thin alpha-disk where most of the radiation is coming from the disk region close to the SMBHs.  For accretion rates $< 0.01 {\dot M}_{\rm Edd}$, the accretion is believed to be operating in the hot accretion mode, where the accretion flow is optically thin and geometrically thick \citep[e.g., Figure~13 in][]{Ho08}.  
For nearby low-luminosity active galactic nuclei (LLAGNs) such as Sgr~A$^*$ and NGC~3115, hot gas density within $R_B$ can be determined through X-ray observations, allowing us to estimate the available material to fuel the supermassive black holes (SMBHs) within their "sphere of influence".  The Bondi accretion rate is given by $\dot{M_B}=4\pi\lambda R^2_B \rho c_s$, where $\rho$ is the gas density near the Bondi radius and $\lambda$ is the adiabatic index of the hot gas \citep{Bon52,Bag+03,WIS+14}.  However, the observed X-ray luminosity of these LLAGNs is orders of magnitude smaller than what would be expected from the Bondi accretion rate with a $\sim$$10\%$ radiative efficiency.  Thus, it is not simply that they are starved for gas to explain such low luminosity.

Extensive theoretical work has been conducted to understand why LLAGNs are under-luminous.  It has been suggested that during the hot accretion process, most of the energy in the gas is carried by the ions and is advected into the black hole before it has a chance to radiate much energy (advection-dominated accretion flows, or ADAFs; \cite{Ich77,RBB+82,NY94}).  It is also possible that matter passing through the Bondi region does not reach the event horizon of the black hole, but either circulates in convective eddies (convective dominated accretion flows or CDAFs; \cite{NYA00,QG00,AIQ+02}, or some of the gas actually escapes the potential of the black hole in an outflow (such as advection-dominated inflow-outflow solutions or ADIOS; \cite{BB99,Beg12}), or variations on these themes.  These models can collectively be classified as radiatively inefficiently accretion flow (RIAF) models.
Numerical hydrodynamic simulations have been conducted to model hot accretion flows, which generally reproduce analytic self-similar solutions during the time-averaged ``steady-state’’ flow in the simulations \citep{Stone1999,Igumenshchev2000}.  These simulations confirm that strong winds or outflows can be driven during accretion \citep{Stone1999, Li2013}.   More realistic magnetohydrodynamic simulations have been conducted, including magnetically arrested disk (MAD) models and other magnetic field-dominated accretion flows that can drive powerful relativistic jets \citep{Igumenshchev2003, Igumenshchev2008, McKinney2012, Tchekhovskoy2012}.  Additionally, the roles of cooling and thermal conduction have been explored within these models \citep{GBT15,Shcherbakov2010,SWI+14}.
A more comprehensive review of these hot accretion models, with more detailed numerical simulations, can be found in \citep{YN14}.

To test and distinguish these RIAF models, one might compare the predictions with the X-ray properties of the hot gas flowing into the Bondi region.  Notably, all of these early hot-accretion models predict a temperature increase towards the center due to the gravitational influence of the black holes.  However, the density slope of hot gas inside the Bondi radius strongly depends on the specific RIAF models.  A direct measurement of the temperature and density profile of the hot X-ray emitting gas both inside and around the Bondi radius is key to understanding and constraining the RIAF models.  Yet, observational constraints have proven to be challenging due to the small angular scale of the Bondi radius of black holes.  Even for the closest SMBHs, the angular sizes of the Bondi radii are on the order of a few arcseconds or less (see Figure~\ref{feasibility} below).  

\begin{figure}[t]
\centering
\includegraphics[width=0.312\textwidth]{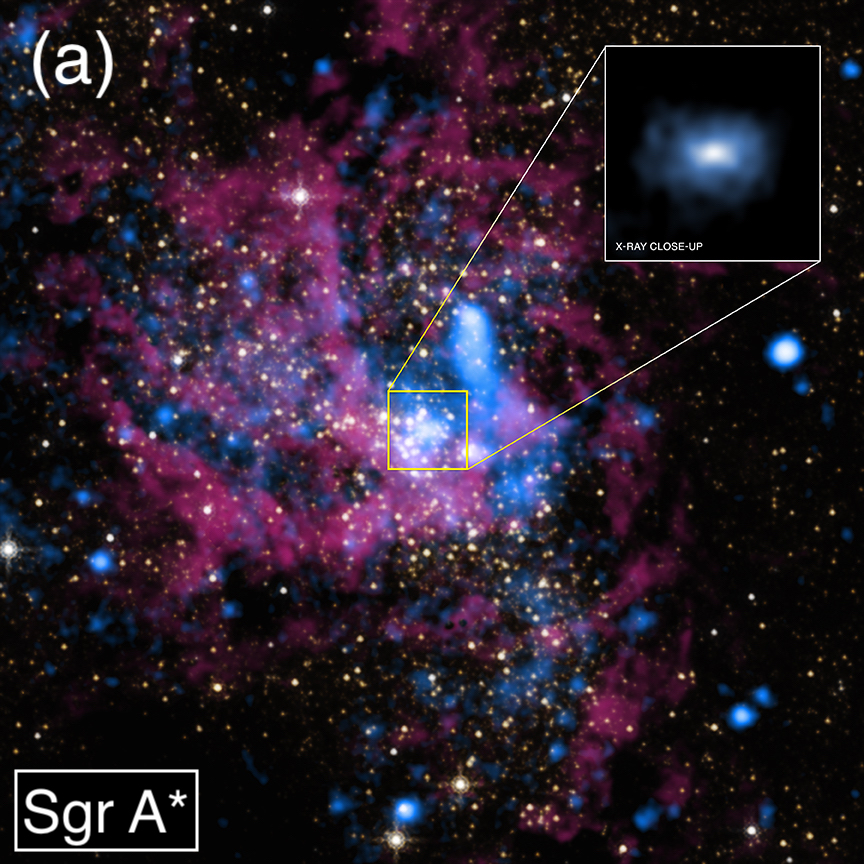}
\includegraphics[width=0.352\textwidth]{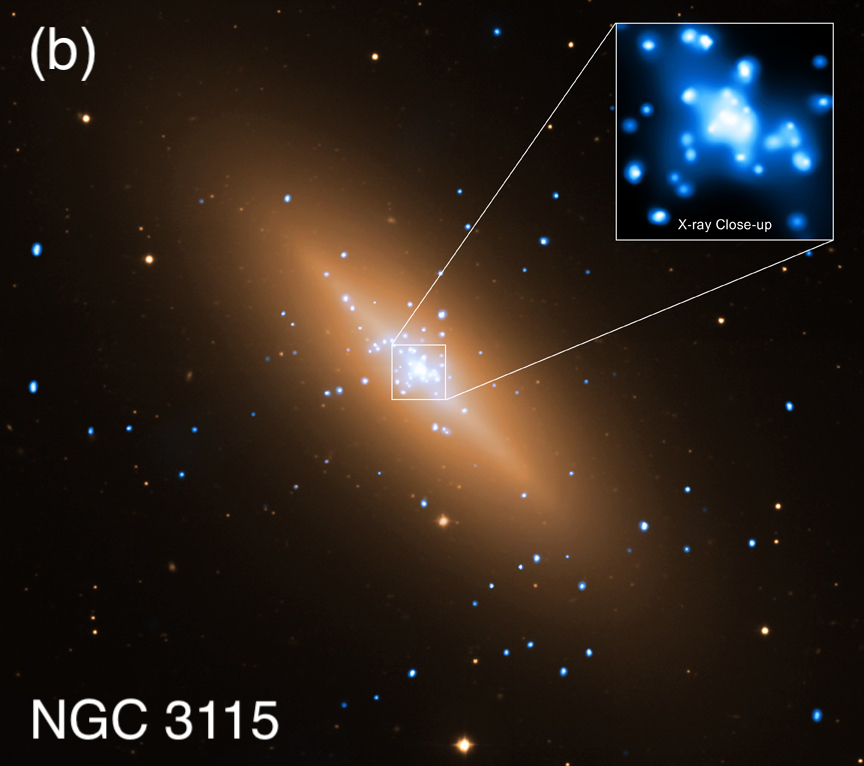}
\includegraphics[width=0.315\textwidth]{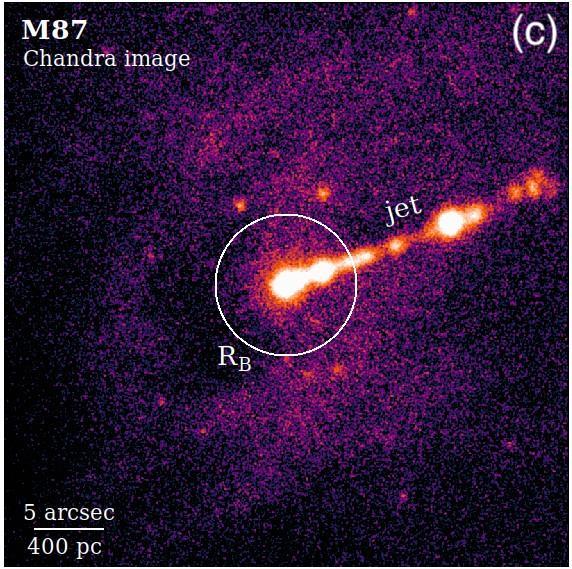}
\includegraphics[width=0.346\textwidth]{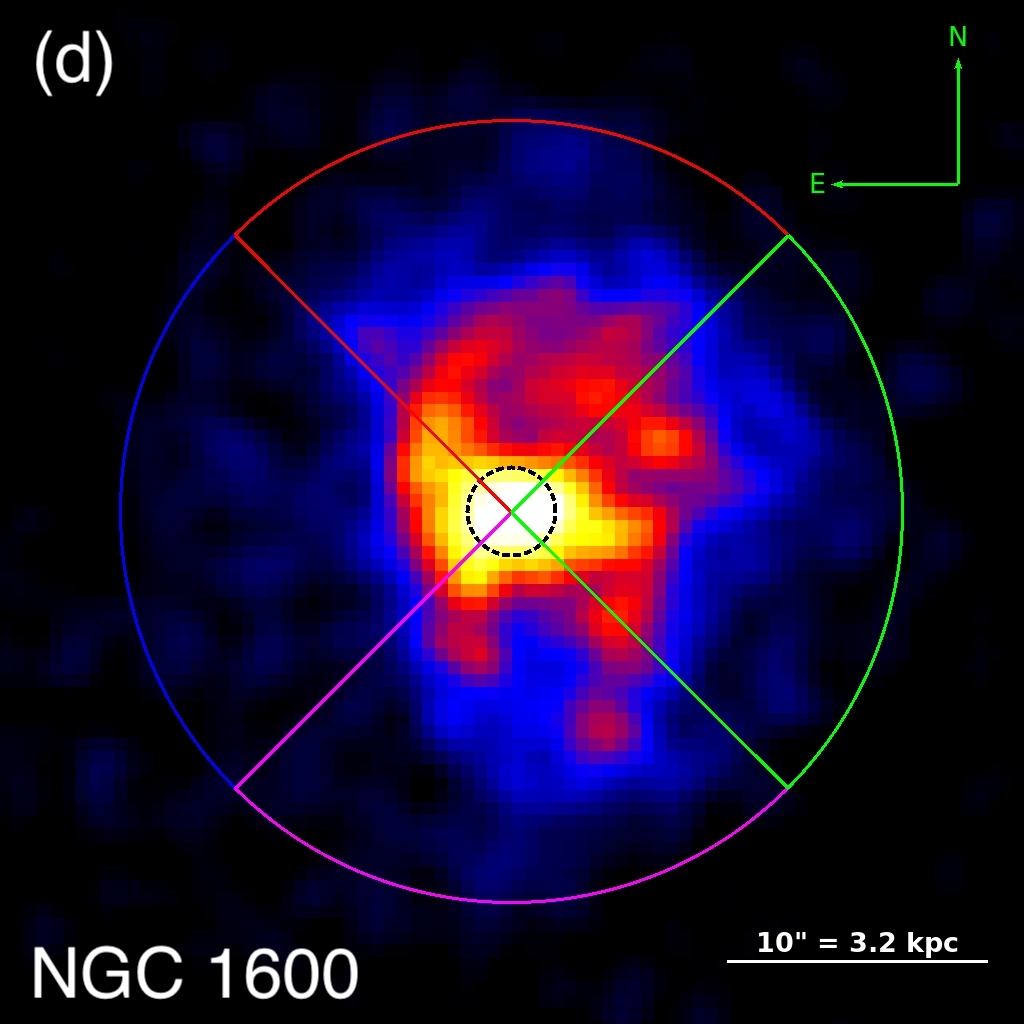}
\includegraphics[width=0.38\textwidth]{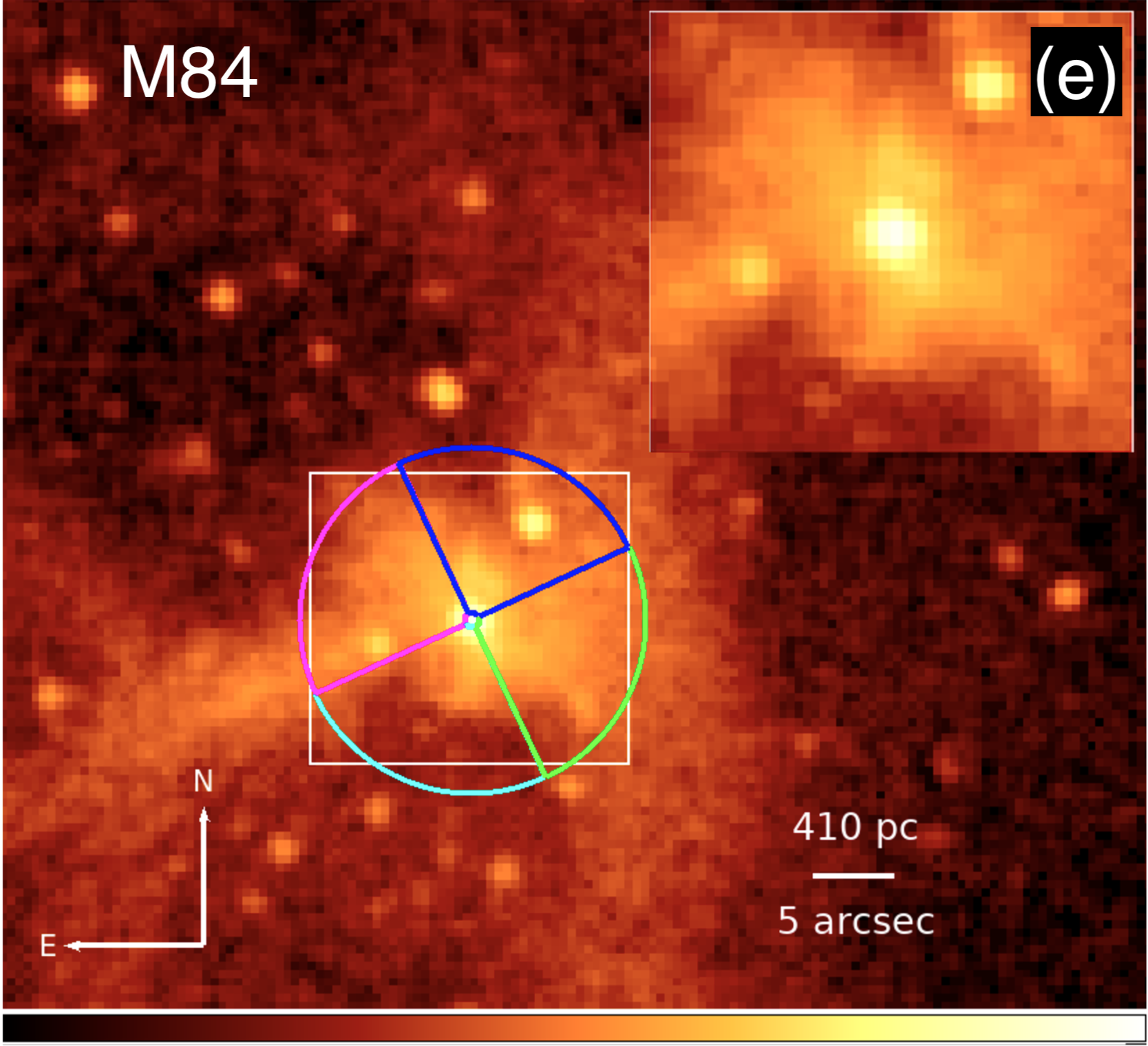}
\caption[Resolved Bondi regions with \textit{Chandra}]{LLAGNs and their host galaxies with Bondi regions spatially resolved with \textit{Chandra}.  From upper left to lower right: (\textbf{a}) Sgr~A$^*$ \citep{Wan+13}, (\textbf{b}) NGC~3115 \citep{WIY+11,WIS+14}, (\textbf{c}) M87 \citep{RFM+15,Rus+18}, (\textbf{d}) NGC~1600 \citep{RW21}, and (\textbf{e}) M84 \citep{BRR+23}.  In the first two panels, X-ray emission is shown in blue, and is also visible in the insets (image credits: NASA/CXC/SAO).  The remaining panels are X-ray images.}
\label{images}
\end{figure}  

\begin{figure}[t]
\centering
\includegraphics[width=4.7 cm]{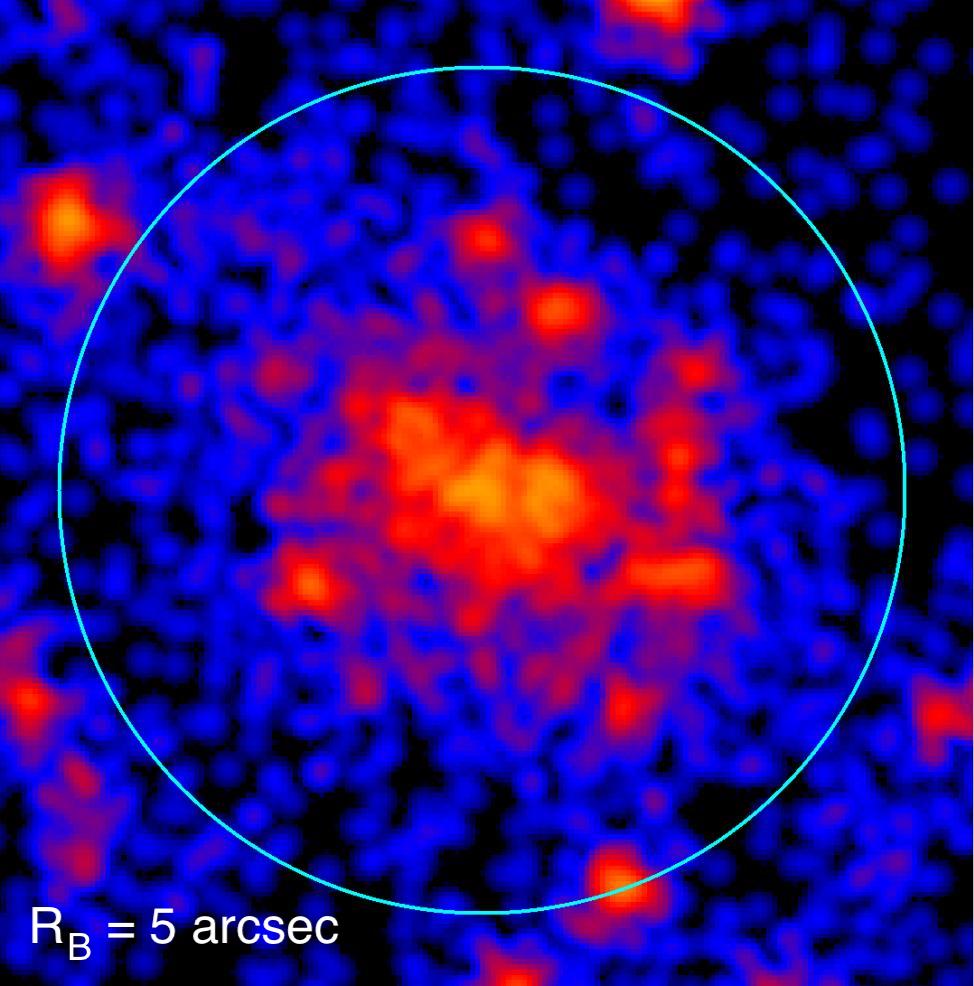}
\includegraphics[width=4.7 cm]{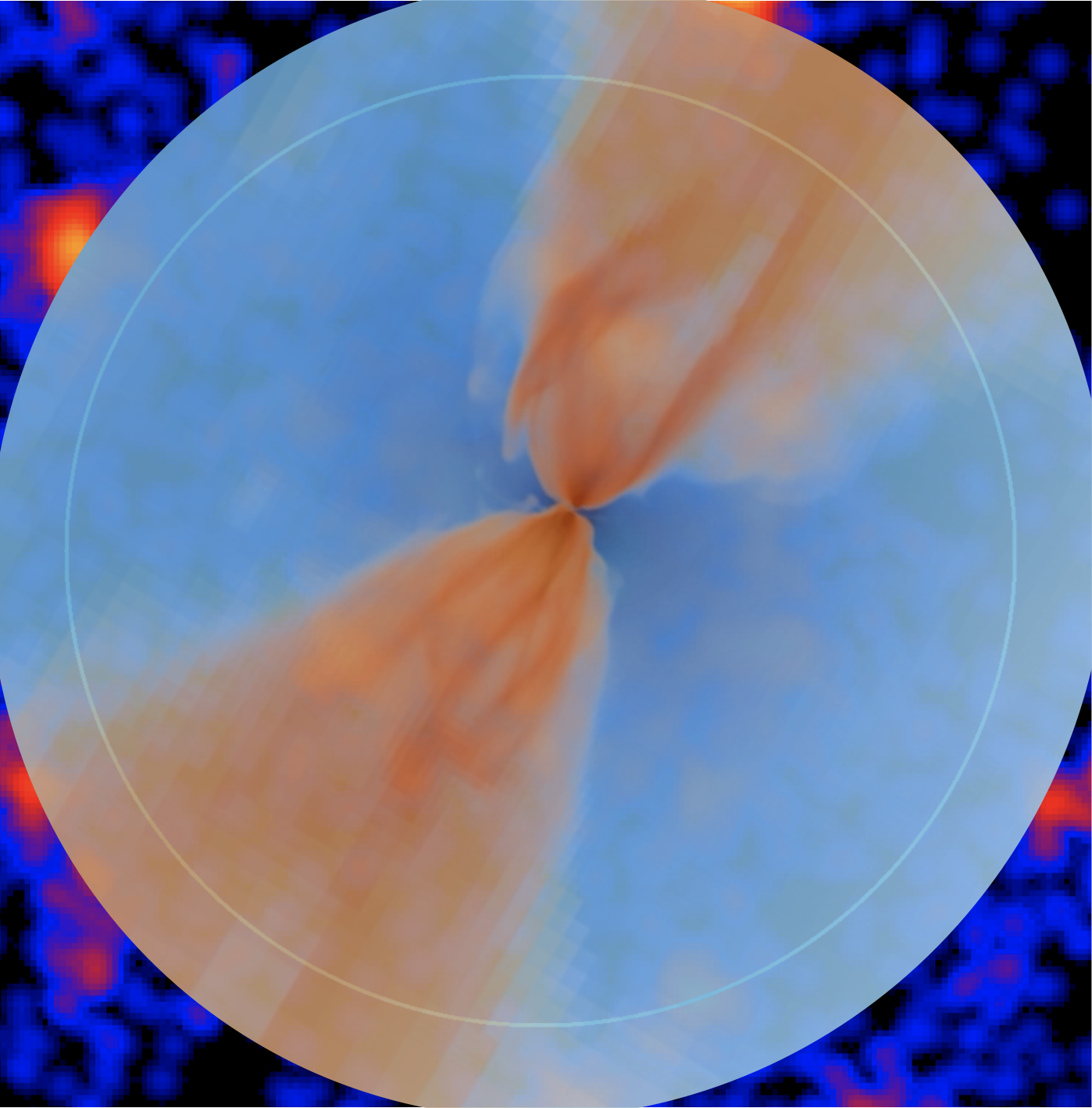}
\includegraphics[width=6.6 cm]{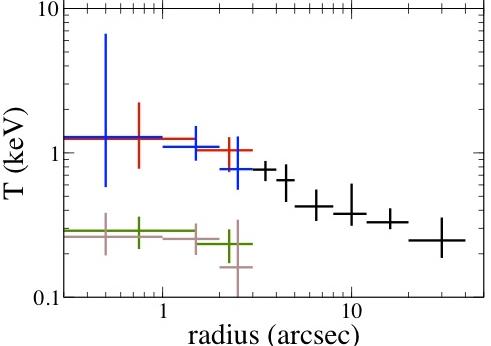}
\caption[Multiphase gas inside Bondi radius]{(\textbf{Left}) A megasecond \textit{Chandra} image of NGC~3115 inside the Bondi radius of about 5 arcsec (circle) \citep{WIS+14}.  Different colors represent different intensity levels, with the brightest region at the center.  (\textbf{Middle}) Multiphase inflow-outflow simulation of a LLAGN overlaid on the \textit{Chandra} image [Qiu et~al. in preparation].  The infalling directions (shown in blue) are cooler, while the outflow directions (shown in orange) are hotter.  The simulation is not specifically tailored to NGC 3115, but is shown here to illustrate the potential of using AXIS temperature maps to test the multiphase model.  (\textbf{Right}) Temperature profiles of NGC~3115 featuring a hotter component of about 1\,keV and a cooler component of about 0.3\,keV inside about 3 arcsec using a two-component model \citep{WIS+14}.  Different colors represent different radial binning schemes.  A single temperature model fit would give a drop in temperature toward the center \citep[see Figure~3 in][]{WIS+14}.}
\label{n3115}
\end{figure} 

\section{Current status in resolving the hot gas within the Bondi radius of SMBHs}
\label{sec:current}

Thanks to the sub-arcsec resolution of \textit{Chandra}, a few of the Bondi regions have finally been resolved and studied in detail \citep{Bag+03,BRR+23,RW21,RFM+15,Rus+18,Wan+13,WIS+14,WIY+11}.  Figure~\ref{images} shows the X-ray images of these Bondi regions; all of these regions belong to LLAGNs and are believed to be undergoing hot accretion.  Although two of these systems were the targets of 
\textit{Chandra} X-ray Visionary Projects (XVP)\footnote{Chandra X-ray Visionary Projects (XVP) were the largest type of science research proposals issued between Cycle 13 and Cycle 16. They involved new Chandra observations for major, coherent science projects aimed at addressing key questions in current astrophysics that required 1 Msec or more of observing time.} 
\citep[Sgr~A$^*$ and NGC~3115,][]{Wan+13,WIS+14,WIY+11}, the angular resolution, spectral capabilities, and photon statistics were just barely adequate for the task, resulting in data that necessitate further study.  In Sgr~A$^*$, the flat density profile of the gas inflow implies the presence of an outflow that expels more than 99\% of the matter initially captured at the Bondi radius \citep{Wan+13}.  NGC~3115 has one of the largest Bondi regions in angular scale, approximately 
2.4--4.8 arcsec \citep{WIS+14,WIY+11}.  Hot gas inside the Bondi region of NGC~3115 has been clearly resolved, providing the sharpest view of the accretion flow to date (Figure~\ref{n3115}).  A flattening of the density profile within the Bondi radius of NGC~3115 also indicates the presence of an outflow.  The measured density profile and inferred outflow have provided important constraints for modeling the full spectral energy distributions (SED) of RIAF modes \citep[e.g.,][]{NSE14,BAF+16,FWL16,ANW+18,MRI+19}.

One of the most surprising results regarding NGC~3115 is the absence of a gas temperature increase towards the center (right panel in Figure~\ref{n3115}).  This finding is in sharp contrast to all the earlier RIAF models.  The temperature inside the Bondi radius appears to be complex, indicative of a multi-temperature phase of hot gas.  More specifically, the X-ray spectrum inside the Bondi radius includes a hotter gas component ($\sim$ 1\,keV) and a cooler one ($\sim$ 0.3\,keV).  The gas may be cooling out of the X-ray hot phase and feeding the rotating gas disk, rather than free falling onto the black hole.  Indeed, more realistic numerical simulations suggest that the large-scale accretion inside the Bondi radius can be more chaotic \citep[e.g.,][]{PSB+17}, with cooler gas raining down in some directions and hotter gas outflowing in the others \citep[e.g.,][]{GRO13,GBT15}.  Additional physical factors such as conduction, stellar feedback, viscosity, and galactic potential have also been investigated to explain the observed temperature and density measurements \citep[e.g.,][]{SWI+14,YG20,SY21,AN20}.

M87 also has one of the largest Bondi radii \citep[$\sim$5 arcsec;][]{RFM+15,Rus+18}.  It is the most gas rich system among those with resolved Bondi radii, providing us with very high photon statistics to study the hot gas spectrum in more detail.  Consistent with the results of NGC~3115, there is also no evidence for the expected temperature increase within the Bondi radius.  This suggests that the hot gas structure is not dictated by the SMBH’s potential and, together with the shallow density profile, indicates that the classical Bondi or ADAF formalism with pure accretion may not be applicable.  In stark contradiction to most theoretical calculations, these observations suggest that the gas flows within the Bondi radius are a complex mixture of inflow fueling the black hole and powerful outflows.

The evidence of hot gas existing in multi-temperature phase is very strong in M87, with the possibility of a mini cooling flow of temperature down to $\sim$0.2\,keV inside the smallest radius resolved by \textit{Chandra} \citep{Rus+18}.  
It is possible that the gas could further cool to lower temperatures but was not detected with \textit{Chandra} due to the limitations of its soft response below about 0.5\,keV, compounded by decreased sensitivity due to ACIS contamination \citep{Plucinsky2022}.
With the high sensitivity of AXIS in soft X-rays \citep{2023_AXIS_Overview}, it will be able to search for even cooler phases of gas within the Bondi radii of M87 and other LLAGNs, which could not be detected with current X-ray observatories.

Recently, the Bondi regions of two additional LLAGNs, NGC~1600 and M84, have been resolved by \textit{Chandra} \citep{BRR+23,RW21}.  The results are very consistent with previous studies.  Notably, neither displays the temperature increase toward the center that is predicted by earlier RIAF models.  Both exhibit fairly flat density profiles, which is more consistent with significant gas outflow.  Multi-temperature hot gas has also been significantly detected in NGC~1600.  However, there is no evidence of multi-temperature gas in M84, perhaps due to its more uniform temperature or the presence of cooler gas below 0.2\,keV that could not be detected with \textit{Chandra}.  In any case, multi-temperature seems to be quite typical inside the Bondi radius of SMBHs.  A more detailed spectral analysis of the multi-temperature phase would require significantly higher photon statistics.

\begin{figure}[t]
\centering
\includegraphics[width=6.62 cm]{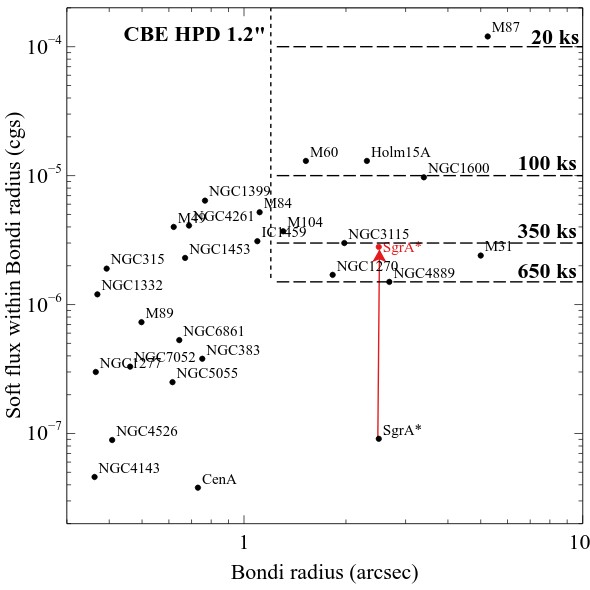}
\hspace{0.5cm}
\includegraphics[width=8.9 cm]{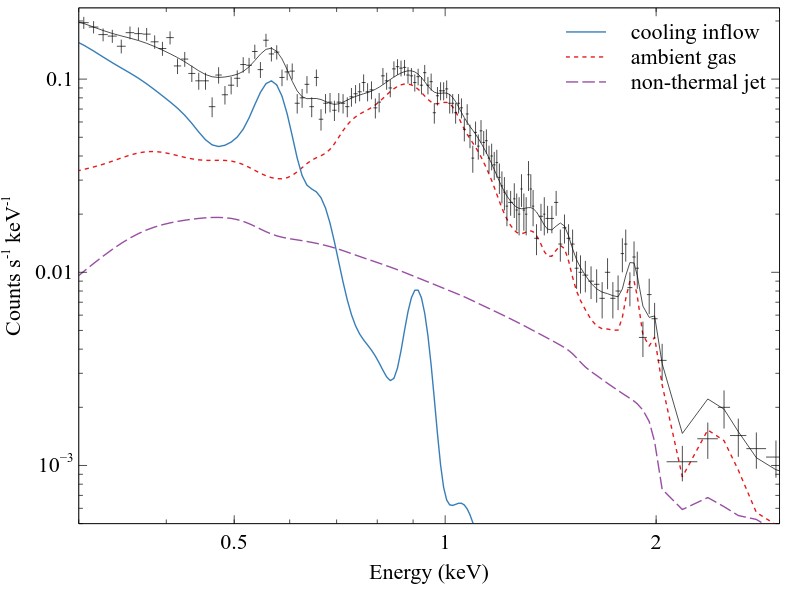}
\caption[Sample of supermassive black holes for AXIS]{(\textbf{Left}) Distribution of the sizes of the Bondi radius\protect\footnotemark for a sample of SMBHs.  The vertical line separates samples that can be resolved with AXIS's current best estimate HPD of 1.2 arcsec, illustrating its ability to measure the Bondi radius in a significant number of sources.  Exposure times required for at least 3000 counts in the soft band 0.5--1\,keV, with the exception of Sgr~A$^*$ where the estimation was based on hard flux (highlighted in red) due to high absorption in soft X-rays.  The total number of counts across the full energy range would be significantly higher.  (\textbf{Right}) Simulated AXIS spectrum of M87 with 100\,ks exposure for a region with the size of AXIS's HPD with a total of $\sim$10,000 photon counts in 0.3--10\,keV (including both thermal gas emission and non-thermal emission from the AGN and jet).
}
\label{feasibility}
\end{figure} 
\footnotetext{It should be noted that the concept of Bondi radius is not well defined in a realistic system due to the multi-temperature nature of the hot gas and the non-trivial outer boundary condition.  Nevertheless, we can still estimate the size of the Bondi radius observationally to characterize the rough size of the accretion region.  However, the sizes shown here are only approximate, as we assume the hot gas temperature to be 0.5\,keV, which underestimates the Bondi radii for cooler systems.} 

\section{Prospects for Bondi radius studies with AXIS}
\label{sec:AXIS}

Although the current studies were somewhat limited by photon statistics, they have supported the idea that LLAGNs are generally associated with strong outflow, and in the most extreme cases, take the form of realistic jets (e.g., M87).  
While the most powerful observed jets from active galaxies are associated with radio quasars accreting in a cold mode near the Eddington limit,
the connection between hot accretion flows and outflows/jets has also been extensively discussed \citep{YN14}.  A more detailed study of hot gas inside and near the Bondi radius of SMBHs, using the finest spatial resolution, will be crucial for understanding both the accretion mechanism and feedback process in these LLAGNs.

With the current baseline spatial resolution of 1.5 arcsec and the current best estimate of 1.2 arcsec (Half Power Diameter, HPD), AXIS will have a resolution very similar to \textit{Chandra} but with a significantly larger effective area for collecting photons.  This will enable us to resolve and map the emerging hot accretion flow within and around the Bondi radius of about 10 nearby AGNs (Figure~\ref{feasibility}).  AXIS will reveal transitions in the inflow that ultimately fuels the AGN and outflows along the jet-axis that limit the accretion rate.

With the estimated exposures listed in Figure~\ref{feasibility}, AXIS will be able to collect at least 3000 counts in the soft band (0.5--1\,keV) within $R_B$, enabling us to measure density and temperature with less than a 10\% uncertainty.  The total number of counts across the full energy range would be significantly higher.  Multi-temperature gas with a cooler component has been detected in three LLAGNs, which has provided important constraints on theoretical models of the accretion flow and feedback models.  The larger sample with higher photon statistics will allow us to search for multi-temperature signature, which appears to be common.  The sample also contains a range of AGN/outflow/jet activities (e.g., from NGC~3115 with no AGN activity, to NGC~1600 with moderate activity, to M87 with a relativistic jet), allowing us to study accretion and outflow in a diverse population of systems and to search for a correlation with jet power.  Note that the sample contains different kinds of LLAGNs, which can have different histories of inflow/outflow near the Bondi radius.  Additional parameters such as velocity dispersion, star formation rate, and black hole mass will be needed to understand the nature of different LLAGNs.

The best targets with the largest Bondi radii, such as NGC~3115 and M87, will be resolved into many more regions, allowing a more detailed comparison with numerical simulations that predict cooler gas in the accretion directions and hotter gas in the outflow directions (e.g., Figure~\ref{n3115}).  
For the brightest target, M87, with a 100\,ks exposure for a region the size of AXIS's HPD inside its Bondi radius, AXIS will collect approximately 10,000 photon counts in the 0.3--10\,keV range, including thermal gas emission and non-thermal emission from the AGN and jet (right panel in Figure~\ref{feasibility}).  In contrast, \textit{Chandra} would only collect about 170 counts in the 0.5--7\,keV range with the same exposure.  This allows for creating a very high spatial resolution temperature map to study the multi-temperature gas inside the Bondi radius of M87 in great detail.
Hot gas around and beyond the Bondi regions will also be mapped with the highest possible resolutions, allowing us to understand the interplay between large-scale accretion and feedback to the environment.  This can also be done for those LLAGNs with Bondi radii just a bit too small to be resolved, significantly increasing the sample size.

For M87 and Sgr~A$^*$, the Event Horizon Telescope (EHT) has resolved radio emission from plasma at event-horizon scales \citep{EHT2019,EHT2022}, providing important constraints on black hole masses and accretion rates down to the gravitational radius, $r_g = GM_{\rm BH}/c^2$, where the gravitational radius is five to six orders of magnitude smaller than the Bondi accretion radius. The process of AGN fueling from the Bondi radius down to the event horizon remains uncertain \citep{Guo2023,Johnson2023}; a holistic view encompassing all scales from the event horizon to the Bondi radius is critical for accurately interpreting the EHT observations \citep{Blandford2022}.  Recently, numerical simulations have been conducted to simulate accretion flows in these systems over a wide range of length scales from the event horizon out to $10^3$--$10^6\,r_g$.  These simulations have revealed a variety of phenomena, including smooth Bondi-like flows, turbulent torus-like structures, shocks, and filaments, and jet launching \citep{Lalakos2022,Olivares2023}.  When cooling is included in the simulations, a multi-temperature gas structure is also observed \citep{Guo2023}, similar to those discussed in Section~\ref{sec:current}.  The Bondi regions observed with AXIS will provide critical boundary conditions for the large-scale accretion flow, linking it to the event-horizon-scale accretion flow in realistic simulations to understand how black holes are fueled.

\section{Concluding Remarks}
\label{sec:conclusion}

We have discussed the current status of results on spatially resolved hot gas within the ``sphere of influence'' of several nearby LLAGNs observed with \textit{Chandra}.  Contrary to earlier models of hot accretion that predicted a steep temperature rise within the Bondi radius, the findings reveal a complex, multi-temperature structure of the hot gas.  The density profiles within these Bondi regions are shallow, indicative of strong outflows.  These observations align with recent numerical simulations that incorporate realistic physics, such as cooling and magnetic fields, suggesting chaotic, multi-temperature accretions with significant outflows.

The gas structure on these scales is key to understanding the following questions:  How are AGNs fueled?  How can an AGN respond to the large-scale cooling rate?  How and why do AGNs evolve from a quasar-era peak to local quiescence?  
To answer these important questions, insights have been obtained from the handful of spatially resolved Bondi regions through deep \textit{Chandra} observations, which suggest strong outflows are expelling most of the gas in all systems, accompanied by rapidly cooling inflow in M87.
The next step is to test whether these processes are sufficient to power the AGN in nearby systems and whether there is a correlation with jet power.  
Currently, the observations are limited by the available photon statistics for spectral analysis.
With the decreased soft response, such studies are now beyond the capabilities of \textit{Chandra} \citep{Plucinsky2022}. 
AXIS would be an excellent choice to carry forward this research across a broader sample of nearby AGNs, mapping the hot gas within and around their ``sphere of influence'' - ranging from inactive systems to those with relativistic jets.  
The science addressed by AXIS also directly ties to the Astro2020 Decadal Survey’s theme of the Cosmic Ecosystem in Stellar and Black Hole Feedback, which seeks to understand ``how the energy from the black hole couples to the surrounding gas'' \citep{NAP26141}.


\vspace{6pt} 



\acknowledgments{We kindly acknowledge the AXIS team for their outstanding scientific and technical work over the past year. This work is the result of several months of discussion in the AXIS-AGN and AXIS-GALAXY SWGs.  KW acknowledges support from SUNY Brockport.}





\section*{References:}
\vspace{-2.5\baselineskip}
\externalbibliography{yes}
\bibliography{references}

\end{document}